\documentclass[conference]{IEEEtran}
\IEEEoverridecommandlockouts

\usepackage{amsmath,amssymb,amsfonts}
\usepackage{graphicx}
\usepackage{textcomp}
\usepackage{import}
\usepackage{booktabs}
\usepackage{comment}
\usepackage{enumerate}
\usepackage{multirow}
\usepackage{authblk}
\usepackage{multirow}
\usepackage{makecell}
\usepackage{lineno}
\usepackage[dvipsnames]{xcolor}

\begin{document}

\title{ Breaking Barriers: Maximizing Array Utilization for Compute In-Memory Fabrics\vspace{-0.25cm}}

\author[1]{Brian Crafton$^\dagger$\thanks{$\dagger$ These authors contributed equally}}
\author[1]{Samuel Spetalnick$^\dagger$}
\author[1]{Gauthaman Murali}
\author[1]{\\Tushar Krishna}
\author[1]{Sung-Kyu Lim}
\author[1]{Arijit Raychowdhury}
\affil[1]{Georgia Institute of Technology, Atlanta, GA}
\affil[1]{School of Electrical and Computer Engineering}
\affil[ ]{brian.crafton@gatech.edu, arijit.raychowdhury@ece.gatech.edu \vspace{-0.5cm}}

\maketitle

\begin{abstract}
Compute in-memory (CIM) is a promising technique that minimizes data transport, the primary performance bottleneck and energy cost of most data intensive applications. This has found wide-spread adoption in accelerating neural networks for machine learning applications.
Utilizing a crossbar architecture with emerging non-volatile memories (eNVM) such as dense resistive random access memory (RRAM) or phase change random access memory (PCRAM), various forms of neural networks can be implemented to greatly reduce power and increase on chip memory capacity. 
However, compute in-memory faces its own limitations at both the circuit and the device levels.
Although compute in-memory using the crossbar architecture can greatly reduce data transport, the rigid nature of these large fixed weight matrices forfeits the flexibility of traditional CMOS and SRAM based designs. 
In this work, we explore the different synchronization barriers that occur from the CIM constraints. 
Furthermore, we propose a new allocation algorithm and data flow based on input data distributions to maximize utilization and performance for compute-in memory based designs. 
We demonstrate a 7.47$\times$ performance improvement over a naive allocation method for CIM accelerators on ResNet18. 
\end{abstract}
\section{Introduction} \label{section:intro}

Modern computing systems are heavily dependent on the capacity and access time of expensive memory banks due to the ever increasing performance gap between main memory and logic. 
Furthermore, the cost of moving data has become more expensive than operating on it \cite{chen2017eyeriss}, and thus not only has the memory become the fundamental bottleneck of computing, but both reading and transporting the data has become more expensive than the operation we seek to perform. 
Popularization of data intensive applications like machine learning and artificial intelligence have further exacerbated this problem.
To address these issues, new architectures based on traditional CMOS attempt to minimize the transport of data by optimizing for data reuse \cite{chen2017eyeriss} and adopting constraints inspired by the brain \cite{davies2018loihi}.
While these techniques yield strong results, they still face the fundamental technological limitations of CMOS. 

Fortunately a new class of embedded non-volatile memory (eNVM) is positioned to minimize data transport by performing compute in-memory. 
In-memory computing seeks to perform matrix multiplication ($\vec{y} = W \vec{x}$) in a crossbar structure using Ohm's law and the non-volatile conductance state provided by the non-volatile memory.
Using this technique, each weight of the matrix ($W_{ij}$) is programmed as a conductance to a cell and each value of the vector ($\vec{x_i}$) is converted to voltage and applied to the rows of the memory crossbar. 
By Ohm's law, the current through each cell is proportional to the product of the programmed conductance ($W_{ij}$) and applied voltage ($\vec{x}_i$). 
By Kirchhoff’s current law (KCL), the resulting currents summed along the columns of the crossbar are proportional to the product of the matrix and vector, ($\vec{y}$).
Under this procedure, the only data transport required for matrix multiplication is the feature vector ($\vec{x}$) from memory and result ($\vec{y}$) to memory. 
Therefore, in-memory computing eliminates the majority of data transfer and thus energy cost of data intensive operations.

Although compute in-memory using the crossbar architecture can greatly reduce data transport, the rigid nature of these large fixed weight matrices forfeits the flexibility of traditional CMOS and SRAM based designs. 
Given that eNVM has high density and unfortunately high write energy compared to traditional SRAM, CIM-based inference-only designs avoid writing to the eNVM cells once programmed.
While this is advantageous for data transport and energy efficiency, it means each CIM processing element (PE) can only perform operations it has the weights for.
This implies that if there is an unbalanced workload where some PEs operations take longer than others, we cannot simply re-allocate these operations to other PEs.
Therefore, we must use synchronization barriers for all PEs so distributed matrix multiplication completes before another is started. 
In contrast, every CMOS and SRAM based PE are computationally identical and can perform any operation in the DNN graph. 

Therefore a fundamental problem in CIM based designs is array utilization, the percent of time an array is in use.
Recent large scale CIM designs \cite{shafiee2016isaac}, use weight duplication and layer pipelining techniques to maximize performance. We describe these techniques in detail in Section \ref{section:background}.
While impressive performance is achieved, these techniques only perform well when the workloads are deterministic. 
Circuit level techniques like zero-skipping greatly increase performance, but create non-deterministic workloads that compromise array utilization. 
In this work we identify and profile these new challenges using a simple simulator framework. 
We then propose a novel algorithm, which makes use of input statistics to find optimal array allocation policies to maximize utilization and \textit{break} synchronization \textit{barriers}. 
Furthermore, we introduce a new data flow that generalizes CIM arrays to maximize their utilization.
We run our experiments on ImageNet using ResNet18 and CIFAR10 using VGG11. 
Although we apply our techniques to deep learning, we claim that the techniques we propose can be extended to any compute in-memory application.
We note that a combination of these strategies yield $7.47\times$ improvement in performance over a baseline naive array allocation. 
\section{Background and Motivation} \label {section:background}

\begin{figure}
\centering
\includegraphics[width=0.48\textwidth]{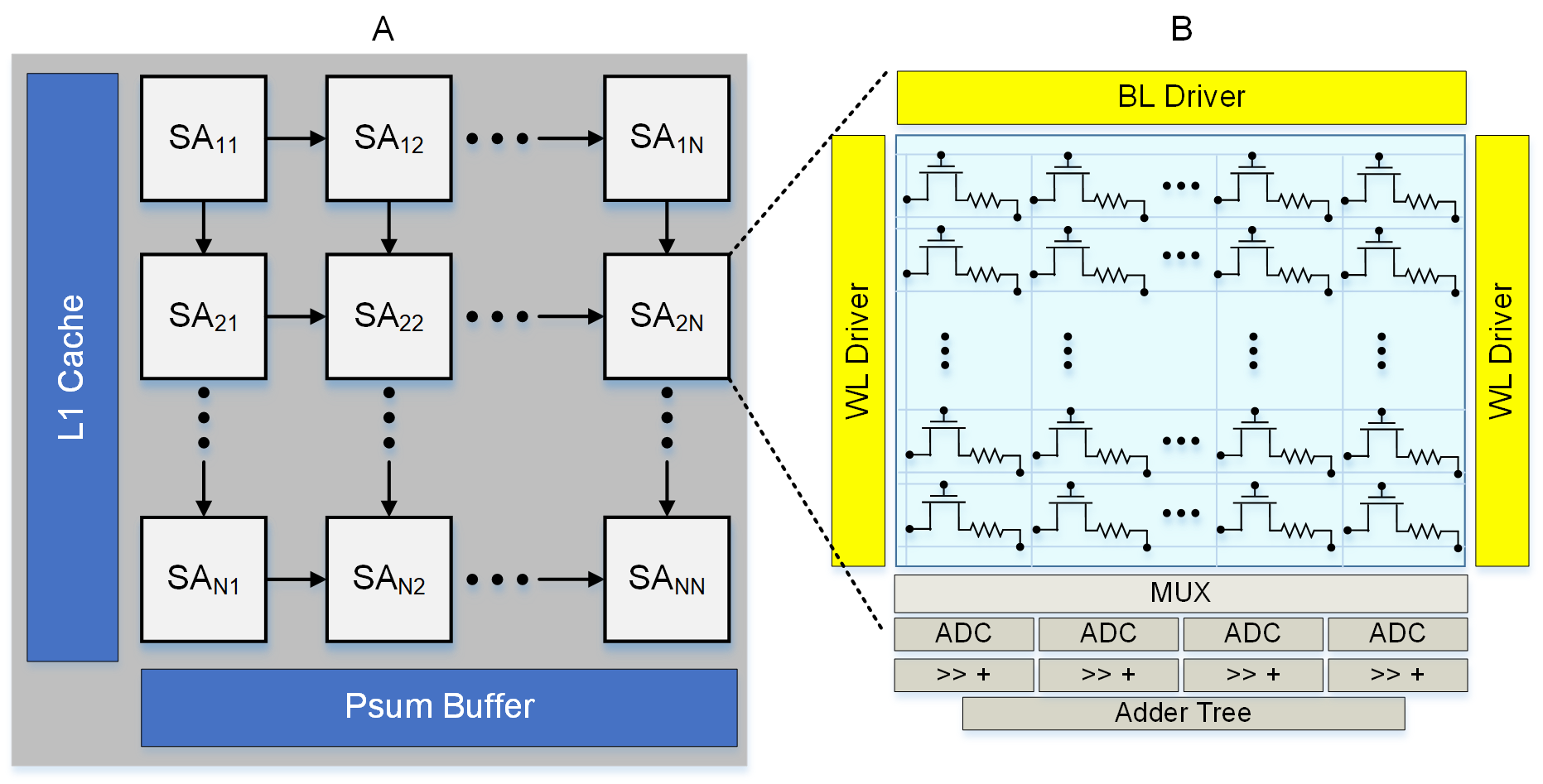}
\vspace{-0.25cm}
\caption{
Typical compute in-memory PE (processing engine) and sub-array (SA) architecture. (A) NxN sub-array PE with L1 cache and psum buffer. In this work N is 8. (B) Typical sub-array design with dual word line drivers, ADCs, shift and add units, and an adder tree.}
\label{fig:arch}
\end{figure}

Compute in-memory systems use binary or multi-level cells as weights to perform matrix multiplication in memory. 
In this work we will focus our attention to binary cells given the current state of the art in eNVM \cite{wu201840nm} already struggles with variance thus making multi-level cells even more difficult to utilize. 
However, the same techniques demonstrated in this work can easily be applied to multi-level cells as well. 
Given binary cells, we must use 8 adjacent cells to form a single 8-bit weight, like those shown in the columns of Figure~\ref{fig:arch}. 
The 8-bit vector inputs to this array are shifted in 1 bit at a time, and the resulting binary product collected at the ADCs is shifted left by the same amount the inputs are shifted right. 
In this way, each array is able to perform an 8-bit matrix multiplication. 

\begin{figure}
\centering
\includegraphics[width=0.33\textwidth]{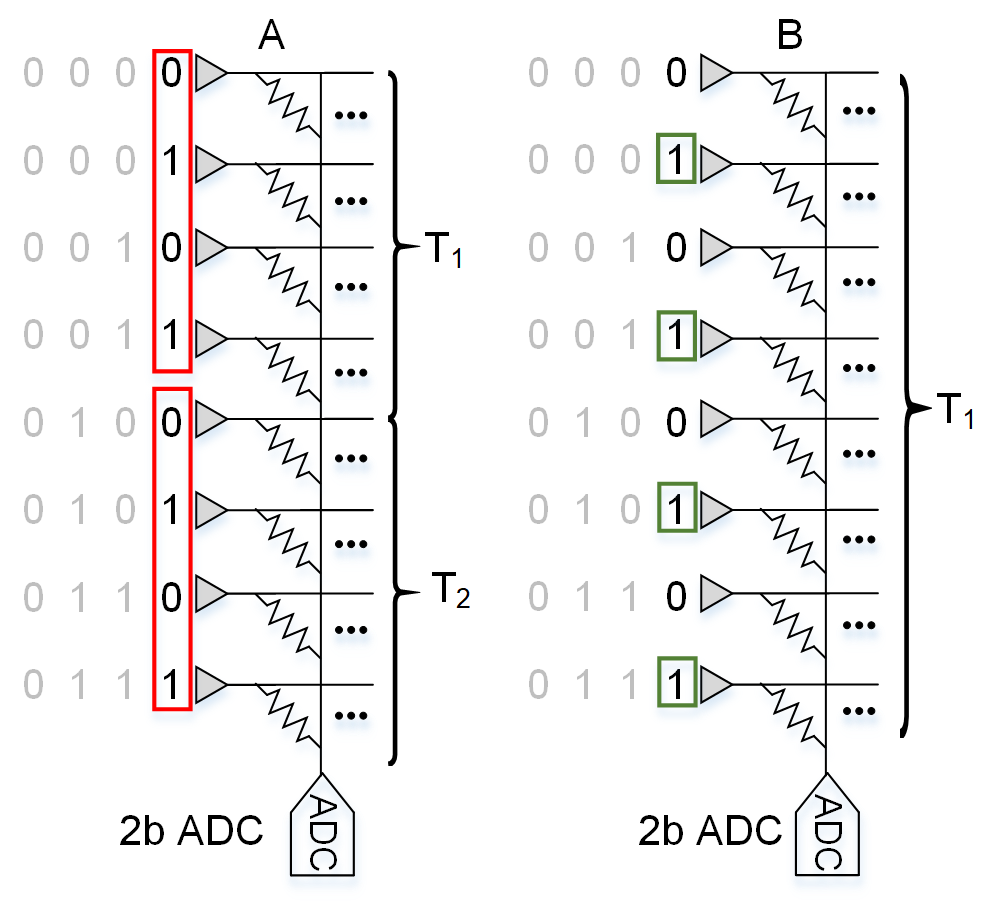}
\vspace{-0.25cm}
\caption{Simplified breakdown of ADC reads in baseline and zero-skipping with 2-bit ADC precision. 
(A) Baseline targets four consecutive rows at a time since the 2-bit ADCs are capable of distinguishing 4 states. 
(B) Zero skipping targets the next 4 rows where the word line is enabled. This way we can read more rows and not overflow our ADC.
}
\label{fig:zero_skip}
\end{figure}

There are two common techniques for performing compute in memory. 
The first technique, we call \textit{baseline}, is simply reading as many rows as the ADC precision allows (e.g. for a 3-bit ADC, we read 8 rows simultaneously). 
The next technique is commonly called zero skipping \cite{yang2019sparse}, where only rows with `1's are read. 
This technique exploits sparsity in the input features or activations (for neural networks). 
Zero skipping performs faster than the baseline technique because for most cases it will process more total rows per cycle.
In Figure \ref{fig:zero_skip}, we provide an example case for zero-skipping where 8 total rows are read using a 2-bit ADC.
Baseline (\ref{fig:zero_skip}A) requires 2 cycles since it targets four consecutive rows at a time. 
Zero-skipping (\ref{fig:zero_skip}B) is able to finish all 8 rows in a single cycle because we only consider the `1's in the input vector. 
There are few reasons not to perform zero skipping, unless there is limited input data bandwidth or the eNVM has high variance and accumulated too many errors. 


\begin{figure}
\centering
\includegraphics[width=0.48\textwidth]{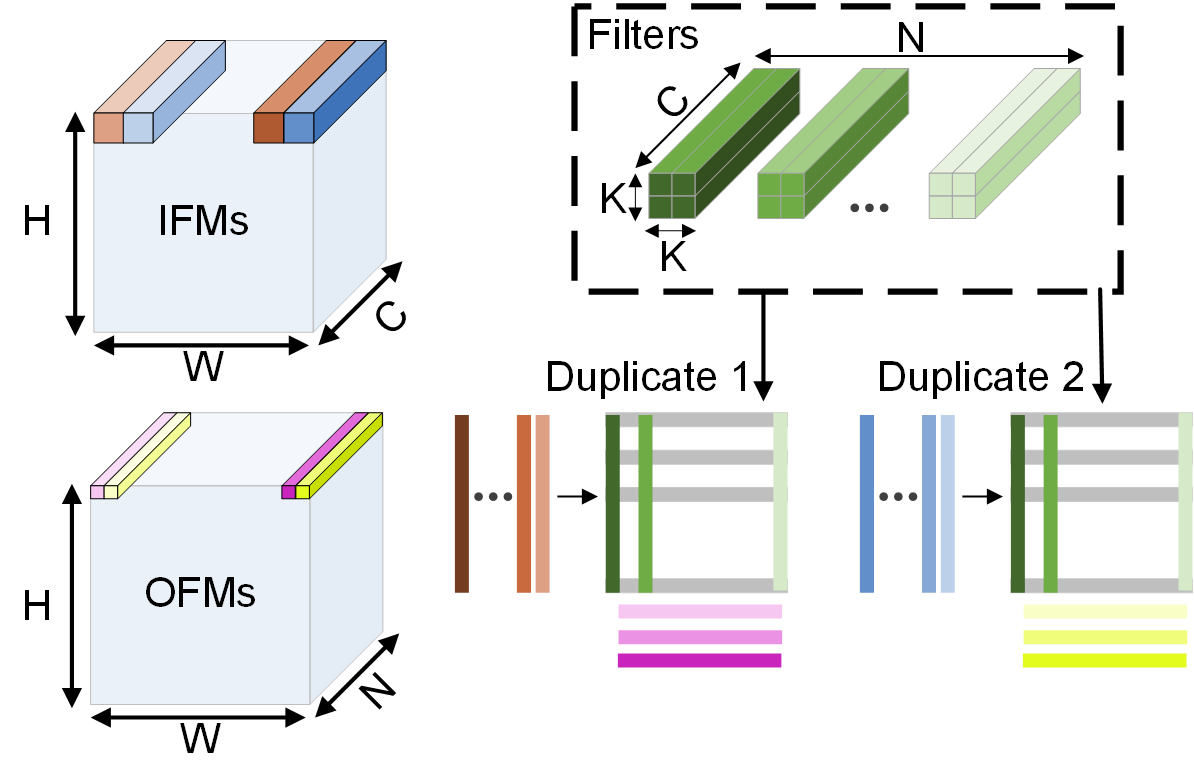}
\vspace{-0.25cm}
\caption{
Convolutional layer mapped to a CIM array. 
Both input features maps (IFM) and filters are vectorized with the filters forming the columns of a matrix.
The vectorized feature maps applied to the crossbar to perform matrix multiplication, where the results are output feature maps (OFMs).
}
\label{fig:cim_conv}
\end{figure}

By encapsulating the array, ADCs, and shift and add logic, a matrix multiplication engine can be created.  
Using these arrays as building blocks, prior work has implemented CNNs (Convolutional Neural Networks) where a group of arrays implement a larger matrix multiplication. 
Despite performing more complex operations, the core operations of CNNs are converted into matrix multiplication. 
In Figure \ref{fig:arch} we illustrate this idea, showing how a group of arrays is tiled together to form a PE. 
In Figure \ref{fig:cim_conv} we further depict how these arrays can be pieced together to form a larger matrix. 
In this example, both input feature maps and filters are vectorized with the filters forming the columns of a matrix.
The vectorized feature maps are input to the crossbar to perform matrix multiplication, where the results are output feature maps for this layer in a CNN.

Given the high density of these PEs, hundreds or thousands of them can be tiled in the same area used by modern ICs. 
Although similar in concept, CIM-based DNN accelerators have numerous differences from traditional CMOS based designs that introduce challenges in maximizing performance. 
First off, a CIM-based PE has fixed weights that cannot be reprogrammed due to the high energy cost of writing eNVM. 
Traditional CMOS based PEs are generalized compute units that can operate on any input data, since they do not contain fixed weights. 
Thus a fundamental issue in CIM-based accelerators is array utilization. 
Several works have addressed this issue introducing ideas such as weight duplication and layer pipelining. 

\textit{Weight duplication} \cite{shafiee2016isaac} 
is used to maximize throughput in large scale CIM accelerators where the amount of on-chip memory exceeds the number of weights in the model. 
In \cite{peng2019optimizing}, 24,960 arrays are used for a total on-chip memory capacity of nearly 104 MB (2b cells), while only using an area of $250mm^2$. Using this enormous on-chip memory capacity, they not only fit ResNet \cite{he2016deep} but duplicate shallow layers up to 32$\times$. 
When weights are duplicated, the input data is divided equally amongst each duplicate array so they can process in parallel. 
We illustrate this idea for a convolutional layer in Figure \ref{fig:cim_conv}. 
The input patches from the input feature maps (IFMs) are divided into groups based on the number of duplicates, and then mapped to each duplicate. 

\textit{Layer pipelining} \cite{shafiee2016isaac} 
is used to maximize throughput in eNVM CIM accelerator, where arrays are not re-programmed due to large amounts of on-chip memory and high write energy.
At the same time, most modern neural networks contain 20 or more layers that must be processed sequentially. 
Given that most designs use $128\times128$ arrays, it becomes infeasible to partition arrays such that they can be used for each layer without being re-programmed. 
This implies that the majority of PEs would sit idle waiting for their layer to be processed. 
To solve this problem, images are pipelined through the network to keep all arrays utilized. Although this compromises single example latency, it maintains maximum throughput. 

\section{Block-wise Array Allocation} \label{section:bb}

In the previous section, we discussed several techniques that are used in CIM accelerators to increase throughput, but each introduces it's own synchronization barrier that limits array level utilization. 
In this work, we identify two of these barriers and propose our solution to mitigate this problem.
The two techniques that create these barriers are weight duplication and layer pipelining. 
In previous work these barriers were not a problem because array performance was deterministic. 
When zero-skipping is introduced, it instigates these barriers because it introduces non-deterministic computation time for each array. 
Zero skipping will only improve the performance of a CIM accelerator because it simply means each array will perform equal to or faster than the baseline algorithm. 
However, since the number of ones in the input vector of the CIM operation follows a random distribution, the amount of time to finish a dot product is non-deterministic. 
This means that several arrays performing a part of a larger matrix multiplication need to be synchronized to the slowest preforming array. 
As the size of the operation (and number of arrays) increases, the more stalls occur. 
In the following section, we explore the implications of zero skipping at the architectural level. 

\subsection{Identifying Synchronization Barriers}

The non-determinism introduced by zero-skipping induces the need for synchronization barriers. 
A synchronization barrier is required when a group arrays are processing a distributed workload and finish at different times, but must be synchronized before starting another task.
The first barrier occurs at the layer level and is a result of using layer pipelining. 
When the arrays are distributed to each layer, we attempt to divide them evenly so that all layers finish at the same time.
If any layer is consistently performing faster than other layers, it will have to stall because layers downstream will not be able to buffer its outputs. 
Previous work \cite{peng2019optimizing} have allocated arrays to layers based on the number of duplicates required such that all layers in the pipeline complete their workload at the same time, and thus sustain full utilization. 

This allocation method works under the assumption that all arrays perform at the same rate and we can choose the number of arrays on chip.
However, as \cite{yang2019sparse} points out neither of these assumptions will hold in a realistic design. 
Prior works \cite{shafiee2016isaac, peng2019optimizing} assume 128 cells can be read at once using 5 and 8 bit ADCs.
Although feasible in theory, we note that such a design will yield very high error given that the state of the art devices have 5\% device-to-device variance \cite{wu201840nm}, and thus at most 8 rows (3-bit) can be read at once.
Such a design also yields very poor memory density since large (5-8 bit) ADCs occupy over $10\times$ the area of eNVM. 
Instead columns must be processed in batches using zero-skipping, where current summation is used for 8 rows and then intermediate results are stored and accumulated using existing digital logic in the array.

\begin{figure}
\includegraphics[width=0.45\textwidth]{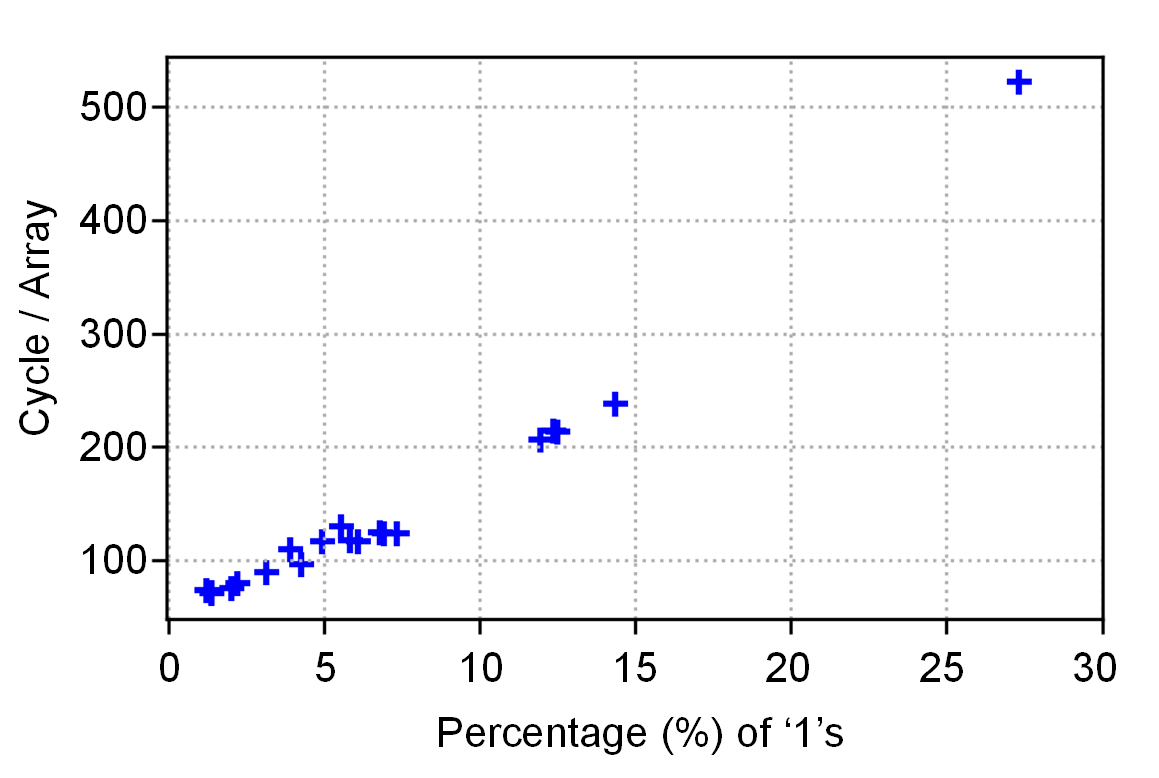}
\vspace{-0.25cm}
\caption{
Cycles per array versus the percentage of `1's in all 8-bit input features. 
Each point represents the average percentage for one of the 20 layers in ResNet18. 
}
\label{fig:dist1}
\end{figure}

When zero skipping is used, each array performs at a non-deterministic speed that follows the distribution of input data it receives.
In Figure \ref{fig:dist1}, we plot the average time for an array to perform a $128\times16$ matrix multiplication versus the percentage of `1's in all the 8-bit input features for the 20 convolutional layers in ResNet18.
To compute the percentage of `1's for a layer, we average the 8 bits in all 8-bit input features together. 
For example, a 1000-entry 8-bit input vector contains 8000 bits and to determine the percentage of '1's, we average over 8000 bits to compute this percentage.
From Figure \ref{fig:dist1}, we infer a linear relationship between the percentage of `1's in the input features to a layer, and the expected number of cycles to perform the matrix multiplication. 

Naturally, we can use this information to better allocate duplicates to each layer in our design. 
We approach this problem by quantifying the total number of multiply-and-accumulate (MAC) operations in each layer, and the average number of MAC operations per cycle an array can perform. 
In prior works, performance per array is constant since each array takes the same number of cycles to perform a matrix multiplication. 
Therefore, arrays are allocated to each layer based only on the total MACs per layer. 
When zero-skipping is introduced and performance per array is not constant, this allocation method fails to allocate evenly.
To achieve equal utilization, we can instead allocate arrays to each layer based on the expected number of cycles it will take to finish without any duplicate arrays.
We can compute the expected number of cycles it will take a layer to finish by dividing the total MACs in a layer by the average performance of each array in the layer. 
We call this allocation method \textit{performance-based} allocation, whereas allocation that assumes all arrays perform evenly is \textit{weight-based} allocation.

\begin{figure}[b]
\includegraphics[width=0.45\textwidth]{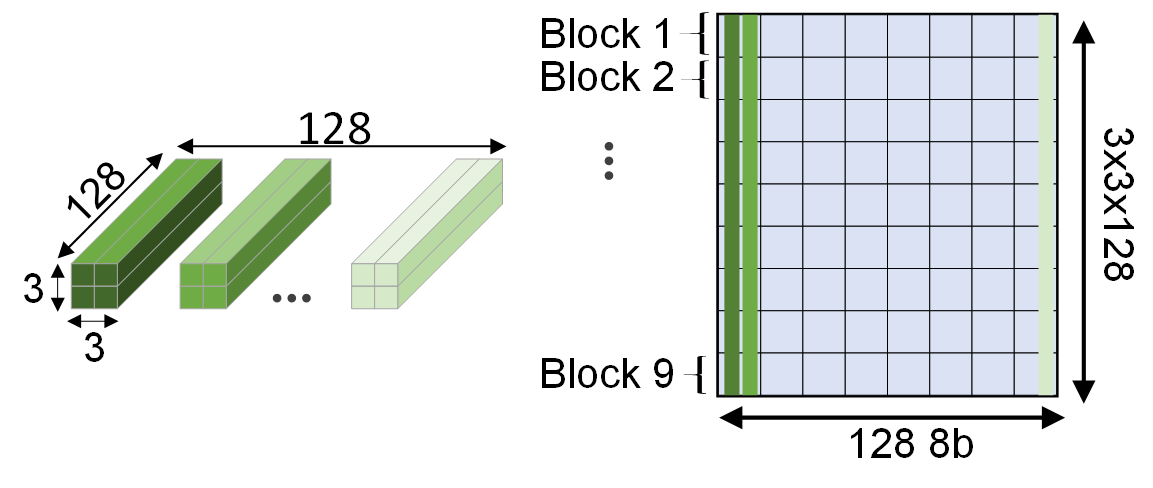}
\vspace{-0.25cm}
\caption{
The $3\times3\times128\times128$ filter used in layer 10 from ResNet18 converted into a matrix with annotated blocks. This filter requires 72 $128\times128$ arrays to store in a $9\times8$ grid. 
}
\label{fig:block}
\end{figure}

\begin{figure}
\includegraphics[width=0.45\textwidth]{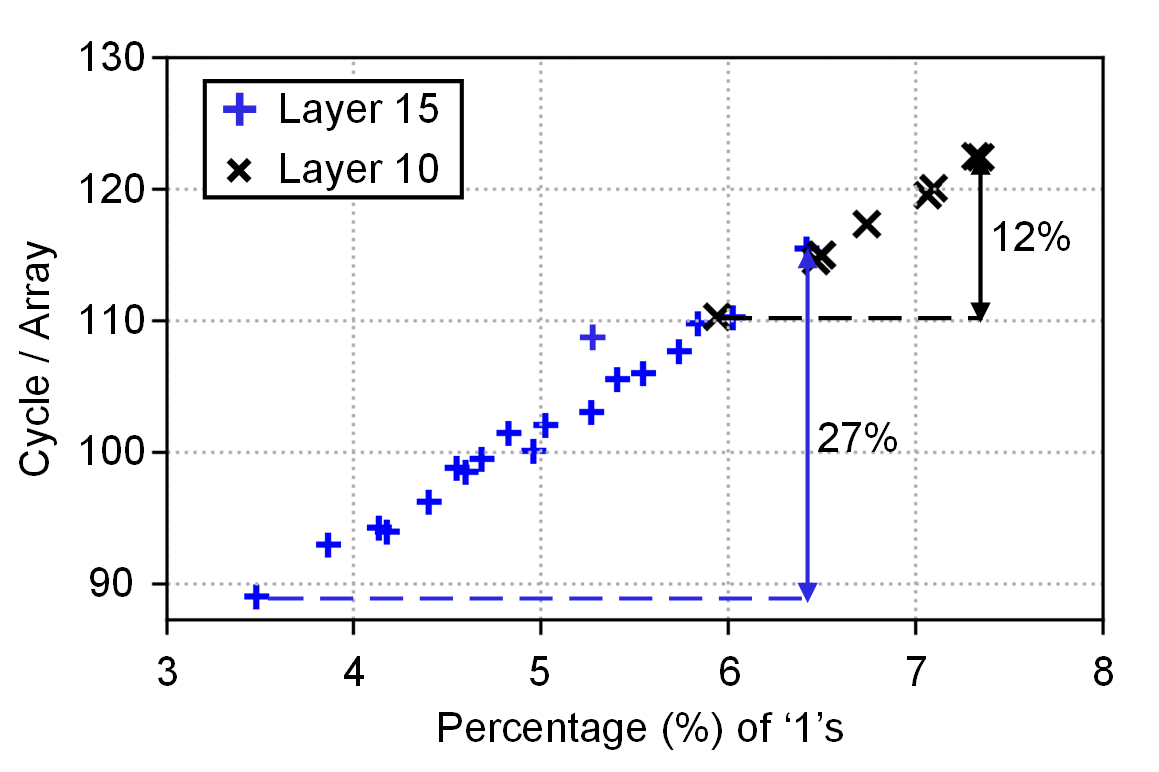}
\vspace{-0.25cm}
\caption{
Cycles per array versus the percentage of `1's in all 8-bit input features. 
The blue crosses represent the average percentage for 1 of the 18 blocks in layer 15 of ResNet18. 
The black $\times$s represent 1 of the 9 blocks in layer 10.
}
\label{fig:dist2}
\end{figure}

While this technique ensures that all our layers will be equally utilized, it does not ensure that the arrays inside each layer will be equally utilized. 
Each layer in our DNN (convolution or fully connected) is implemented as a matrix consisting of eNVM arrays. 
We visualize this idea in Figure \ref{fig:block}, where a $3\times3\times128\times128$ filter is mapped to 72 arrays arranged in a $9\times8$ grid.
In each of the 9 rows, all 8 arrays share the same input data and, consequently, the same word lines. 
This implies that all 8 arrays will operate at the same speed and form our minimal deterministic compute unit that we call a \textit{block}. 
Because the 9 different rows do not share the same input vectors, they will operate at different speeds. 
If some arrays receive fewer `1's than other arrays, they will sit idle waiting for arrays that receive more `1's to finish. 
In Figure \ref{fig:dist2}, we plot the average cycle time of the arrays in each block of layers 10 and 15 (ResNet18) versus the percent of `1's they receive.
Layer 10 is a $3\times3\times128\times128$ filter (Figure \ref{fig:block}) that contains 9 different blocks, and Layer 15 is a $3\times3\times256\times256$ filter that contains 18 different blocks.
Just as before, we observe a linear relationship between cycle time and the percentage of `1's.
Since layer 15 contains more blocks, it is more susceptible to longer delays because the expected slowest block's cycle time increases with the number of arrays. 
In this figure, we observe a 12\% and 27\% difference in cycle time for layers 10 and 15, which motivates a better allocation technique to prevent significant idle time. 

\subsection{Optimizing Array Allocation}

Finding the optimal allocation policy for blocks is more difficult. 
We cannot add redundant blocks to the same layer, because each layer only uses each weight once per operation. 
Instead, we adopt a new grouping strategy for arrays: rather than duplicating layers of arrays, we duplicate blocks of arrays. 
To find the optimal array allocation policy, we propose a linear time ($O(N)$ complexity) solution. 
This is especially important for larger networks like ResNet18, where there are 247 blocks and finding an optimal solution could be quite difficult. 

With this new grouping strategy, we can allocate using the same technique as before.
First we gather an approximation of the average MAC per cycle for each block of arrays.
We can do this two ways. 
The first option, is running a cycle accurate simulator on some example data to get a very accurate approximation.
The second option is to profile the distribution of `1's in the activations gathered from a large set of examples run on a GPU.
Once we have an approximation for the MAC per cycle of each block, we can compute the expected number of cycles each block will take to perform it's partial dot product.
Once we have cycle approximations for each block, we begin allocating arrays to each block. 
While we have free (not allocated) arrays, we loop through and allocate arrays to the block with the highest expected latency.
Once we run out of arrays or the number of arrays left over is not enough to allocate to the slowest block we have found the optimal allocation. 
We call this allocation method \textit{block-wise}, whereas allocation based on the layer is \textit{layer-wise}.

\subsection{Block-wise Data Flow}

To make use of our new allocation policy, a new data flow strategy is required. 
Since arrays from the same layer are not grouped together, we treat these blocks as generalized compute units rather than being bound to a specific duplicate. 
Therefore, we no longer stall for the slowest block in a layer, but rather just send work to the next available block. 
This means that the same blocks will no longer be working together on the same input data, and thus will not be part of the same gather and accumulate procedure. 
As a result, a new routing and scheduling policy is required because blocks will not always send their partial sums to the same accumulator for every input feature map.
To implement this idea, we include output feature destination addresses in the packet containing data when sending input features to each block.
Upon completing a partial dot product, a block sends their computed partial sums to the designated accumulator and requests additional work from the memory controller.
\section{CIM-based Architecture} \label {section:architecture}

Although our allocation policy will work for any general CIM based accelerator, we adopted a similar architecture to previous work \cite{shafiee2016isaac, peng2019optimizing}.
Our basic processing element (PE) contains 64 128$\times$128 arrays. 
We choose 64 arrays because it provides each block with sufficient network bandwidth and SRAM capacity, while maintaining good SRAM density and low interconnect overhead. 
Our input data, weights, and activations are all 8 bits.
Each array has 1 3-bit ADC for every 8 columns where a single column is pitch-matched with a comparator. 
We choose 3-bit because state of the art devices \cite{wu201840nm} have 5\% variance and 3-bits is the maximum precision that can be read with no error.
We shift one bit from each of the 128 inputs in one at a time which takes 8 cycles. 
In the best case scenario, we perform all 128 rows at the same time. In the worst case scenario, it takes 16 cycles since we enable every single row. 
Therefore, each array takes anywhere from 64 to 1024 cycles and performs a 128$\times$16 dot product. 
In all designs we consider, we use use the same 64 array PE and simply increase the count per design. 


\begin{figure}[b]
\centering
\includegraphics[width=0.45\textwidth]{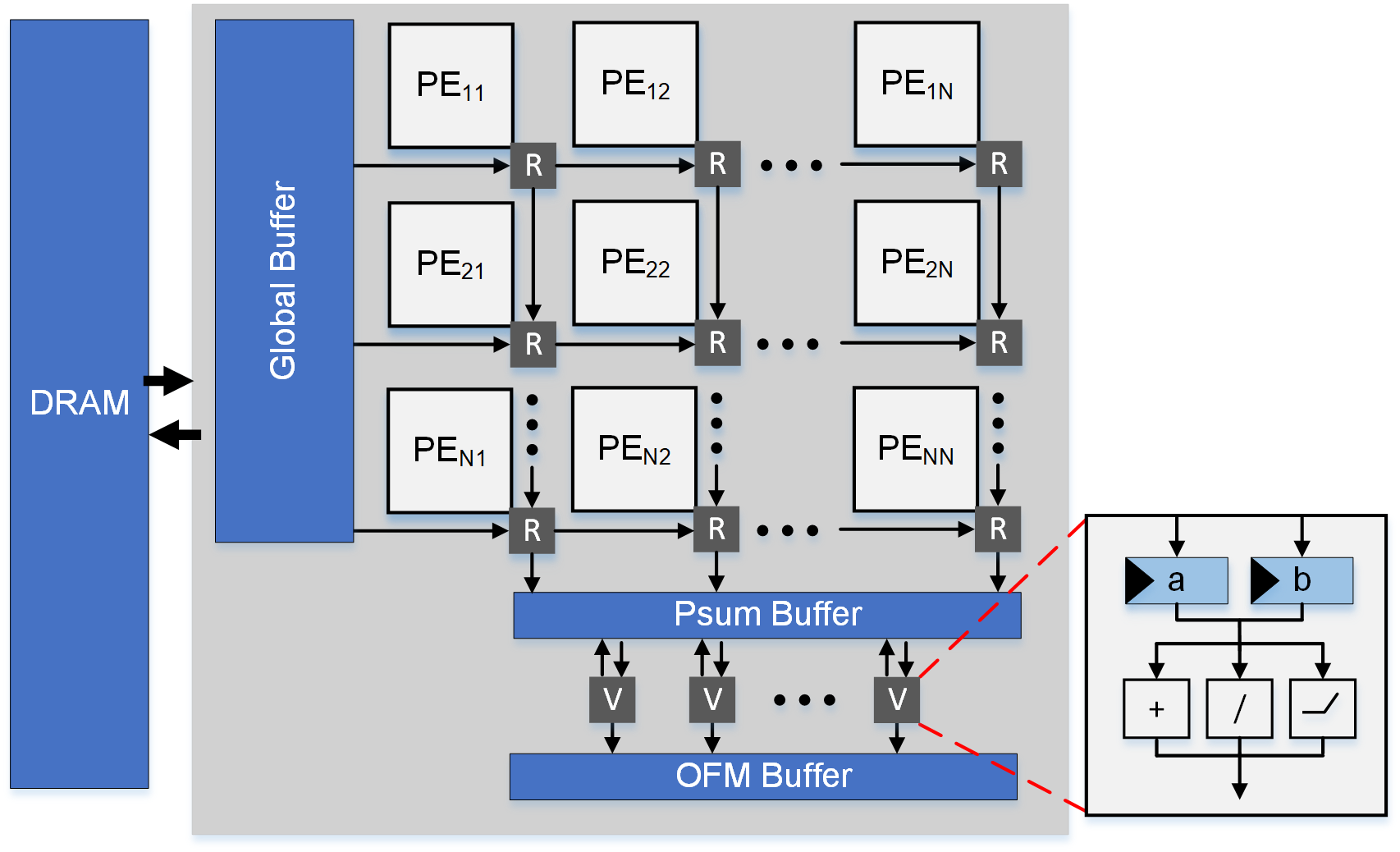}
\caption{
Block-wise network architecture with 1 router (R) per PE. 
All input features are routed from the global buffer to PEs. 
All partial sums are routed from PE to vector unit (V), and vector unit to output feature buffer. 
}
\label{fig:network}
\end{figure}

The activation inputs to the RRAM sub-arrays are stored in on-chip SRAM, while the input images are read in from external DRAM.  
Matrix multiplication is performed by the PEs, while custom vector units are used to perform vector-wise accumulation, bias addition, quantization, and relu.
We use a $N \times N$ mesh network for communication between PEs, memory, and vector units shown in Figure \ref{fig:network}. 
Since blocks vary in size and no block contains 64 sub-arrays, we have to partition the PE to contain several blocks. 
This configuration implies that the different blocks share the same virtualized input and output ports. 
As discussed in Section \ref{section:bb}, input and output vectors are packetized to include destination information. 
Each block in the PE is given an id that is used to route packets to and from. 
Upon completing a partial dot product, a block sends its partial sum to vector units where they are accumulated and activation functions and quantization is applied. 
\section{Results} \label {section:results}

\begin{figure}
\centering
\includegraphics[width=0.45\textwidth]{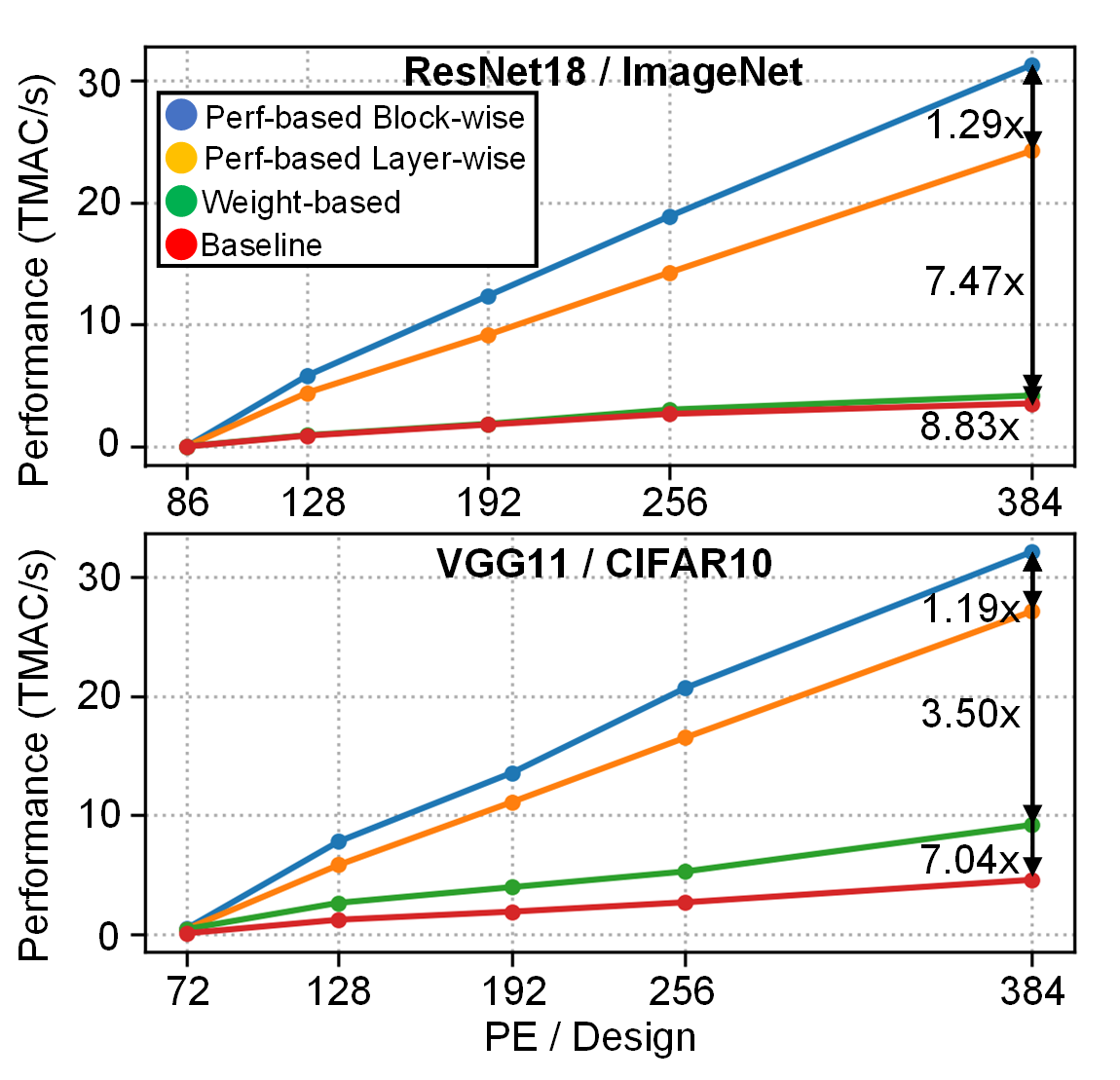}
\vspace{-0.25cm}
\caption{ 
Inference performance for ResNet18 and VGG11 by algorithm and design size assuming 100MHz clock. 
For ResNet18, block-wise allocation sustains a 8.83$\times$, 7.47$\times$, and 1.29$\times$ speedup over baseline (no zero-skipping), weight-based, and performance-based layer-wise allocation.
For VGG11, block-wise allocation sustains a 7.04$\times$, 3.50$\times$, and 1.19$\times$ speedup.
}
\label{fig:results1}
\end{figure}

\begin{figure*}
\centering
\includegraphics[width=0.95\textwidth]{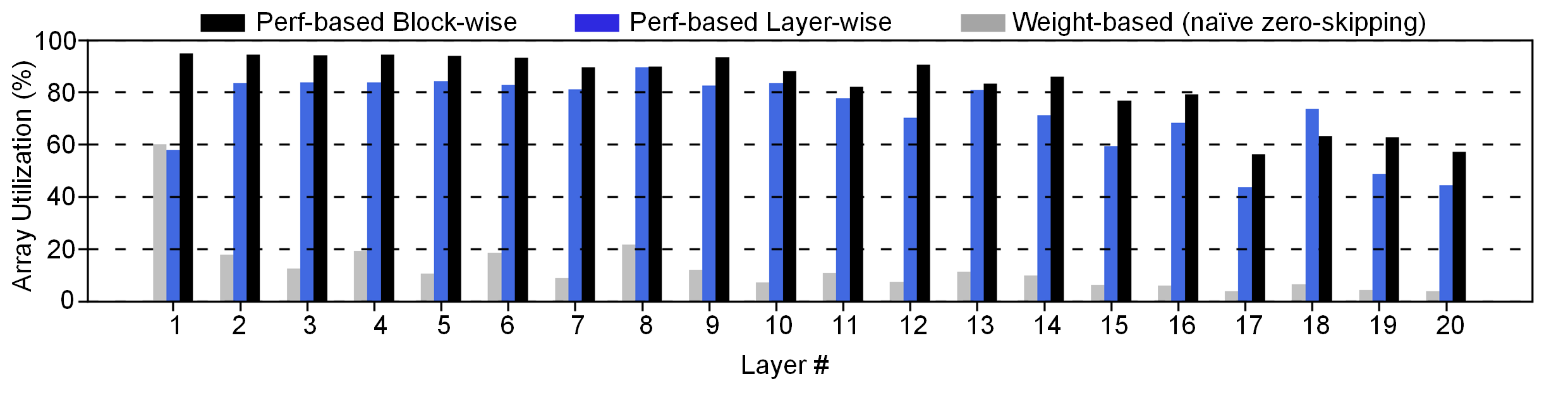}
\vspace{-0.4cm}
\caption{
Array utilization by layer for ResNet18 on ImageNet. Baseline not shown because zero skipping is not used. 
}
\vspace{-0.5cm}
\label{fig:results2}
\end{figure*}

To benchmark block-wise allocation, we compare with several other techniques: weight-based allocation, performance-based layer-wise allocation, and the baseline algorithm which does not use zero-skipping.  
We empirically evaluate performance and array utilization for the three techniques on ImageNet using ResNet18 and CIFAR10 using VGG11. 
We run these techniques in a custom simulation framework designed to evaluate performance and power of compute in-memory using standard CMOS and RRAM models from \cite{chen2018neurosim}.
In this work we focus on performance evaluations, however higher array utilization will result in less leakage power and improved energy efficiency.

Our simulator performs cycle-accurate implementations of convolutional and fully connected layers. 
It is based in Python, but runs array level operations in C for faster evaluation. 
We model components in the design in object oriented fashion, iterating through all components in all PEs each cycle.
We embed performance counters in our ADC and sub-array objects to track metrics like stalls so we can calculate utilization. 
As input, the simulator takes the network weights, input images, PE level configuration, and chip-level configuration. 
The PE-level configuration includes details like the precision of eac ADC and size of the sub-array. The chip-level configuration contains the number of PEs and details about array allocation and mapping. 
As output, the simulator produces a table with all desired performance counters and all intermediate layer activations that are verified against a TensorFlow implementation for correctness. 

To show how our algorithm scales by the size of the design, we have evaluated the different allocation algorithms on several different designs with increasing numbers of PEs. 
In Figure \ref{fig:results1}, we plot performance versus the number of PEs in the design for both ResNet18 and VGG11. 
For ResNet18, we begin at 86 PEs since this contains the minimum number of arrays (5472) required to store ResNet18. 
At 86 PEs, all algorithms yield the same result since no duplication can be done and weights are simply allocated to store ResNet18. 
From there, we begin increasing the design size by $\frac{1}{2}$ powers of 2. 
Block-wise allocation performs the best achieving 29\% improvement over layerwise-allocation and $7.47\times$ improvement over both weight-based and baseline (not zero-skipping) algorithms. 
We follow the same procedure for VGG11, however we observe that block-wise allocation yields less performance advantage. 
This is because VGG11 has roughly half the layers that ResNet18 has. 
It is more difficult to allocate evenly amongst a deeper network and therefore, block-wise allocation yields better results on deeper networks.

To better understand why we get these large performance improvements, it is useful to analyze array utilization. 
In Figure \ref{fig:results2}, we visualize layer-wise utilization of the 20 convolutional layers from ResNet18 using the different techniques.
It is clear that block-wise allocation sustains the highest array utilization across nearly all layers in the network, easily outperforming the other techniques.
Weight-based allocation performs very poorly because of the very different speeds of each layer and block we showed in Figures \ref{fig:dist1} and \ref{fig:dist2}.
It should be noted that we do not plot the baseline algorithm because it has different array level performance given that zero skipping is not used. 
\section{Conclusion} \label {section:conclusion}

In this paper we demonstrate the efficacy of a new technique and data flow to improve array utilization in CIM accelerators. 
Given that the write energy of eNVM is high, CIM arrays contain fixed weights unlike CMOS PEs which can perform any operation in a DNN.
Thus array utilization becomes a key challenge since only some arrays can perform particular operations.
By profiling input statistics and relaxing our data flow, we can allocate arrays to maximize utilization and as a result, performance.
The proposed allocation algorithm and data flow performs 7.47$\times$ better than naive allocation and a layer-wise dataflow. 

\section{Acknowledgement} \label {acknowledgement}
\noindent
This work was funded by the U.S. Department of Defense’s Multidisciplinary University Research Initiatives (MURI) Program under grant number FOA: N00014-16-R-FO05 
and the Semiconductor Research Corporation under the Center for Brain Inspired Computing (C-BRIC) 
and Qualcomm. 
\small{
\bibliographystyle{ieeetr}
\bibliography{main}
}

\end{document}